\documentclass[12pt,preprint]{aastex}

\def\mj{M$_{\rm J}\ $}
\def\rj{R$_{\rm J}\ $}
\def\etal{{et~al.\,}}
\def\mo{M$_\odot$}
\def\ro{R$_\odot$}

\def\mp{M$_{\rm p}$}
\def\rp{R$_{\rm p}\,$}

\def\mstar{M$_{\ast}$}
\def\rstar{R$_{\ast}$}
\def\teff{T$_{\rm eff}\,$}

\def\mic{$\mu$m$\,$}

\def\sles{\lower2pt\hbox{$\buildrel {\scriptstyle <}
   \over {\scriptstyle\sim}$}}
\def\sgreat{\lower2pt\hbox{$\buildrel {\scriptstyle >}
   \over {\scriptstyle\sim}$}}

\begin{document}

%\slugcomment{Submitted to Ap.J.}

\title{Theoretical Interpretation of the Measurements of the Secondary Eclipses of TrES-1 and HD209458b}

\author{A. Burrows\altaffilmark{1}, I. Hubeny\altaffilmark{1}, \& D. Sudarsky\altaffilmark{1}} 

\altaffiltext{1}{Department of Astronomy and Steward Observatory, 
                 The University of Arizona, Tucson, AZ \ 85721;
                 burrows@zenith.as.arizona.edu, hubeny@aegis.as.arizona.edu, sudarsky@as.arizona.edu}

\begin{abstract}

We calculate the planet-star flux-density ratios as a function
of wavelength from 0.5 \mic to 25 \mic for the transiting extrasolar giant planets
TrES-1 and HD209458b and compare them with the recent Spitzer/IRAC-MIPS
secondary eclipse data in the 4.5, 8.0, and 24 \mic bands.  With only three data points
and generic calibration issues, detailed conclusions are 
difficult, but inferences regarding atmospheric
composition, temperature, and global circulation can be made.  Our 
models reproduce the observations reasonably well, but not perfectly,
and we speculate on the theoretical consequences of variations around our baseline models. 
One preliminary conclusion is that we may be seeing in the data indications 
that the day side of a close-in extrasolar giant planet is brighter in
the mid-infrared than its night side, unlike Jupiter and Saturn.
This correspondence will be further tested when the data anticipated 
in other Spitzer bands are acquired, and we make predictions for what 
those data may show.

\end{abstract}

\keywords{stars: individual (TrES-1, HD209458)---(stars:) planetary systems---planets and satellites: general}

\section{Introduction}
\label{intro}

High-precision radial-velocity measurements of many nearby stellar
primaries have revealed the presence of 
extrasolar giant planets (EGPs) in orbit around them (Mayor \& Queloz 1995; Marcy \& Butler 1996; 
Marcy \& Butler 1998; Marcy, Cochran, \& Major 2000, and references therein)
\footnote{see J. Schneider's Extrasolar Planet Encyclopaedia at http://www.obspm.fr/encycl/encycl.html
and the Carnegie/California compilation at http://exoplanets.org
for a more-or-less up-to-date database of extrasolar planets and their primaries.}.
From these data, the projected mass (\mp$\sin(i)$, where \mp\ is 
the planet mass and $i$ is the orbital inclination) and orbital parameters
of more than 150 planetary companions have been determined.  These data
have revolutionized the study of planetary systems, not the least because their orbits and masses
vary widely and are in general quite unlike those of our Jovian and ice giants.

\begin{figure}
\epsscale{1.00}
\vspace*{-0.7in}
\plotone{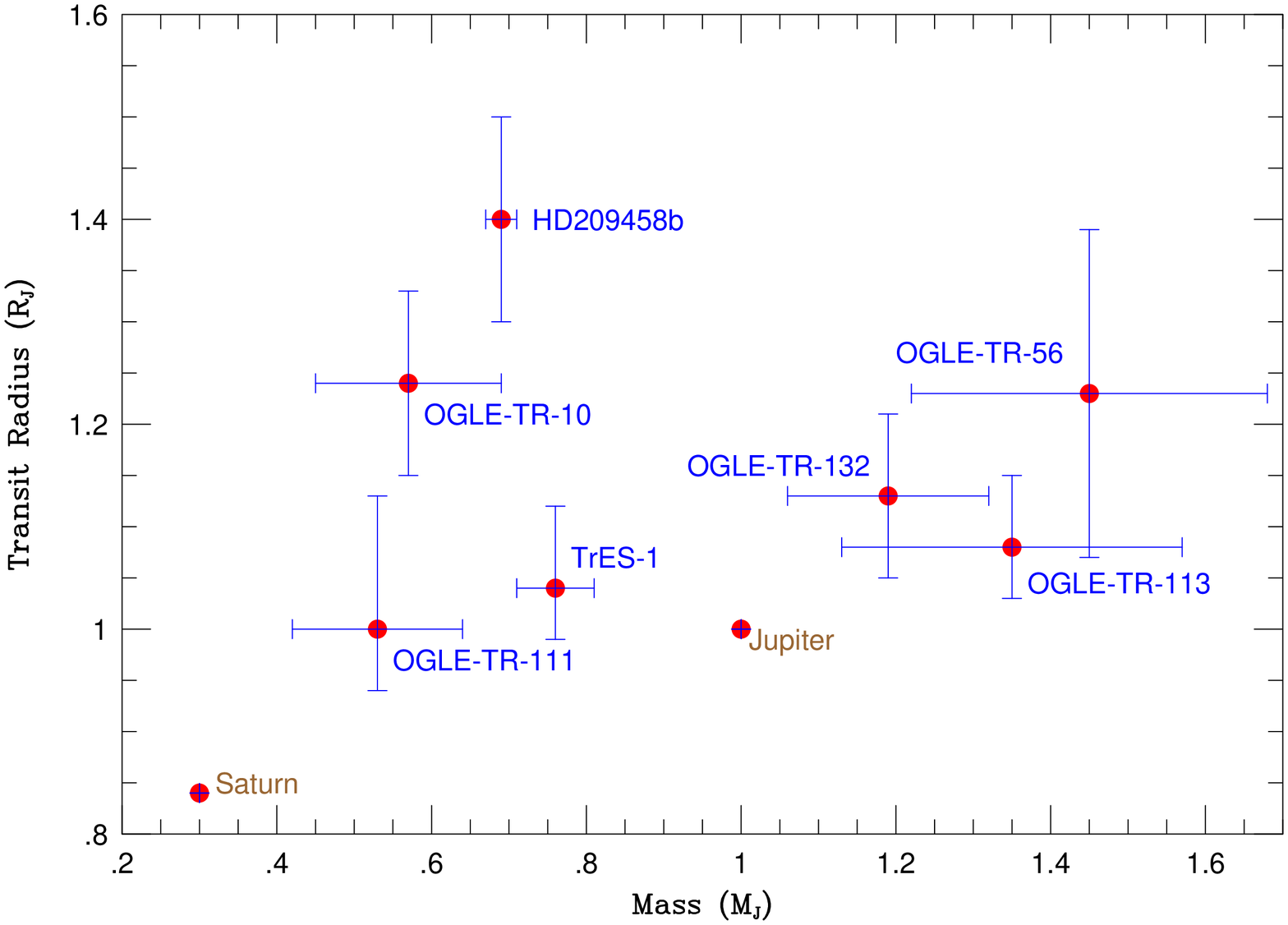}
\vspace*{-0.2in}
\caption{
Transit radii (in units of Jupiter's radius) with error bars 
versus planet mass with error bars (in units of Jupiter's
mass) for the seven EGPs currently seen to transit their primaries.
The positions of Jupiter and Saturn themselves are included for comparison.
Note that HD209458b is the largest transiting EGP (Mazeh \etal 2000; Brown \etal 2001; Cody
and Sasselov 2002) and, as such, is an outlier, though most of these irradiated
EGPs are clearly larger than Jupiter.  An extended transit radius is a known consequence of stellar irradiation
(Guillot et al. 1996; Burrows \etal 2000; Burrows, Sudarsky, \& Hubbard 2003; 
Baraffe \etal 2003; Chabrier \etal 2004; Burrows et al. 2004).
\label{fig:1}}
\end{figure}

The detection of a few transiting EGPs (currently seven), whose orbits are nearly edge-on,
has provided radii, masses, and inclinations (Henry \etal 2000; Charbonneau \etal 2000;
Brown \etal 2001; Sasselov 2003; Konacki \etal 2003ab; Konacki \etal 2004; 
Bouchy \etal 2004; Pont et al. 2004; Torres \etal 2005).
For HD209458b, there are in addition indications of its atmospheric
composition from the wavelength-dependence of the photometric dip of the stellar
light during transit (Charbonneau et al. 2002; Vidal-Madjar et al. 2003; \S\ref{data}).
Figure \ref{fig:1} depicts the mass-radius plot for these transiting EGPs,
with Jupiter and Saturn themselves superposed.

However, it is only by direct detection of the planets and their 
photometric and spectroscopic characterization that they can be
studied in depth to reveal their atmospheric and physical properties (Sudarsky, Burrows,
\& Hubeny 2003; Allard et al. 2003; Burrows, Sudarsky, \& Hubeny 2004; Burrows 2005).
While orbital distances ($a$), periods ($P$), and eccentricities ($e$) 
are reasonably well determined, the investigation of an EGP in physical detail
requires at a minimum the actual mass (\mp), radius (\rp), and composition.
Space provides the needed access, but it had generally been thought that 
high-contrast imaging is necessary to separate out the planet from the bright star.
However, Charbonneau et al. (2005) and Deming et al. (2005) have recently 
shown that variations in the summed light of the planet and star for the
close-in transiting EGPs TrES-1 and HD209458b can be detected with Spitzer
during secondary eclipse.  Secondary eclipse is $\sim$180$^{\circ}$ out
of phase with the transit and is when 
the planet is occulted by the star, thereby shutting off the 
planet's contribution to the summed light.  The approximate magnitude of this
diminution varies significantly with wavelength, with the mid- to far-infrared
being the most favorable bands. At 10-30 microns and in the Spitzer/IRAC
bands from 3.6 \mic to 8.0 \mic this contrast was predicted to be $\sim 10^{-3}$ (Burrows, 
Sudarsky, \& Hubeny 2003; Burrows 2005).  This is near what has now
been detected.  In this paper, we customize to 
the TrES-1 and HD209458 systems the calculation of the planet-star
flux-density ratio as a function of wavelength and compare 
the resulting theory with the Spitzer secondary eclipse data to 
draw conclusions about close-in EGP atmospheres. In \S\ref{data}, we summarize
the new Spitzer data and in \S\ref{model} we briefly describe our numerical
techniques.  Then, in \S\ref{comparison} we present our results from 1 \mic to 25 \mic
and wrap up in \S\ref{discuss} with a discussion of our conclusions
and the outstanding issues concerning irradiated EGPs that such eclipse data 
might address.

\section{Summary of TrES-1 and HD209458b Data}
\label{data}

The parameters of a transiting planetary system 
determine the inputs to a theoretical calculation of its
spectrum and its planet-star flux-density ratio as a function of wavelength.  
Incident radiation at the planet's surface is a function
of the stellar flux density and the orbital distance, and the planet's mass 
and radius determine the gravity of its atmosphere.  For both TrES-1 and
HD209458b, the interior flux (Burrows et al. 2000) is dwarfed by the irradiation effects for
any reasonable system ages.  Since we are not performing evolutionary
calculations (Burrows et al. 2000; Burrows, Sudarsky, \& Hubbard 2003 (BSH); Baraffe et al. 2003) 
in this paper, the system ages are not germane to the problem at hand,
which is reproducing the observed planet-star flux-density ratios in the Spitzer bands.
These ratios depend only on the stellar spectrum, the orbital distance, and
the planet's mass and radius, but not directly on age.

The K0V stellar primary of the transiting extrasolar giant planet TrES-1
is $157\pm 6$ parsecs distant, has a \teff of
$5214\pm 23$ K, a metallicity near solar ($[Fe/H] \sim 0$), a radius (R$_{\ast}$) of $0.83\pm 0.03$ R$_{\odot}$,
a mass (\mstar) of $0.87\pm 0.05$ M$_{\odot}$, and a bolometric luminosity (L$_{\ast}$) near half solar
(Laughlin et al. 2005).  The planet's orbital and physical parameters are a semi-major
axis ($a$) of 0.0393 AU, a period ($P$) of 3.030 days, a planet mass 
(M$_{\rm p}$) of $0.729\pm 0.036$ \mj\footnote{\mj is the mass of Jupiter, $\sim$$1.89914\times 10^{30}$ gm}
(Laughlin et al. 2005) or $0.76\pm 0.05$ \mj (Sozzetti et al. 2004), and a transit radius
(BSH; Burrows et al. 2004) of approximately $1.08\pm 0.05$ \rj 
\footnote{\rj = $7.149\times 10^4$ km, Jupiter's radius} (Alonso et al. 2004; 
Sozzetti et al. 2004; Laughlin et al. 2005).

The corresponding quantities for the F8V/G0V star HD209458 are 47.3 parsecs (Perryman 1997), 
\teff$\sim$6000 K, $[Fe/H] \sim 0$,
R$_{\ast}$ = $1.2\pm 0.1$ \ro, \mstar = $1.1\pm0.1$ \mo, and L$_{\ast}$ = 1.6 L$_{\odot}$
(Henry \etal 2000; Charbonneau \etal 2000; Brown \etal 2001). 
The planet's parameters are $a = 0.0468$ AU, $P = 3.524738$ days, M$_{\rm p}$ = $0.69\pm 0.02$ \mj,
and R$_{\rm p}$ = 1.32 to 1.40 \rj (Henry \etal 2000; Charbonneau \etal 2000; 
Mazeh \etal 2000; Brown \etal 2001; Cody and Sasselov 2002; 
Fortney et al. 2003; Laughlin et al. 2005), with the lower value slightly more favored.
Both TrES-1 and HD209458b have orbital inclinations near 90$^\circ$ and eccentricities near 0.

We see that the flux at the surface of TrES-1 is less than half that at the surface
of HD209458b.  Furthermore, the larger radius of HD209458b implies that it intercepts
more than two times the radiative power.  The former implies that the atmospheric
temperature of HD209458b is likely to be higher than that of TrES-1, while the latter
implies that HD209458b is bolometrically brighter.  Nevertheless, as we show,
their theoretical planet-star ratios are not expected to be very different.

Prior to the recent measurements of the secondary eclipses that motivate this paper,
Charbonneau \etal (2002) had uncovered evidence for the presence of sodium 
in the atmosphere of HD209458b from the different transit radii 
in and out of the Na-D line (see also Fortney et al. 2003).
Similarly, Vidal-Madjar et al. (2003) had seen evidence for atomic hydrogen 
in a stellar-flux-induced planetary wind from HD209458b.  The transit dip in Lyman-$\alpha$
is of such a magnitude ($\sim$15\%) that the planetary material clearly extends
beyond the Roche lobe, indicating planetary mass loss (Burrows 
\& Lunine 1995; Lecavalier des Etangs et al. 2004).  Other than these data,
there had been no determinations to date of the composition of an extrasolar planet.

The new secondary eclipse data for TrES-1 (Charbonneau et al. 2005) and HD209458b (Deming et al. 2005)
hint that this situation may be changing.  TrES-1 shows eclipse depths (planet-star
flux-density ratios) in the Spitzer/IRAC band centered at 4.5 \mic of 0.00066$\pm$0.00013 and 
in the Spitzer/IRAC band centered at 8.0 \mic of 0.00225$\pm$0.00036 (Charbonneau et al. 2005).  HD209458b
shows a corresponding ratio of 0.0026$\pm$0.00045 in the Spitzer/MIPS band centered near 24 \mic 
(Deming et al. 2005).  These numbers are actually the ratios of detected electrons,  
an approximate substitute for the ratio of average flux 
densities in a given band.  For flux density
comparisons, one relies on flux calibrations that may not yet be
robust, particularly given the significant differences between the 
spectra of a close-in EGP and a calibration star (e.g., Vega).
Given this, in \S\ref{comparison} we also provide theoretical 
bandpass-averaged detected-electron ratios.

Since the data from Charbonneau et al. (2005) 
and Deming et al. (2005) are but three of the ten  
potential data points (five bands $\times$ two nearby transiting planets) we can expect 
using Spitzer, very-low-resolution spectra (but ``spectra" 
nevertheless) of EGPs are anticipated soon that will provide compositional and atmospheric
information of an unprecedented character.  
%
%As we show below, using newly-generated theoretical spectra 
%of irradiated EGPs tailored to TrES-1 and HD209458b, we see hints in the 
%secondary eclipse data of the presence of CO and H$_2$O, strong indications
%that these close-in EGP atmospheres are hot, and a weak indication that
%the day side of a close-in EGP radiates more of the absorbed heat 
%than its night side.

\section{Numerical Techniques, Databases, and Assumptions}
\label{model}

The numerical tools we employ to derive the close-in planet's
spectrum during secondary eclipse are described in BSH
and Burrows, Sudarsky, \& Hubeny (2004),
to which we refer the reader for further details.
The spectral/atmosphere code COOLTLUSTY (Hubeny 1988; Hubeny and Lanz 1995; Sudarsky, Burrows,
\& Hubeny 2003) handles the effects of external irradiation using a first-order variant
of the DFE (Discontinuous-Finite-Element) method.  The incident flux is isotropically spread over the
hemisphere of a given planar patch of planetary ``surface."  
As described in BSH, to account in approximate fashion for the 
variation in incident flux with latitude when using a 
planar atmosphere code, as well as the possible day-night 
differences, we introduce the flux parameter $f$.  
A value of $f = 0.5$ assumes that there is little sharing 
of heat between the day and night sides of the EGP.
A value of $f = 0.25$ assumes in the calculation of the $T/P$ profile
that the heat from irradiation is uniformly distributed by efficient 
winds over the entire sphere and that the infrared emissions 
are isotropic.  For this study of the planet-star 
flux-density ratios of close-in EGPs, we use a fiducial value for $f$  
of 0.25, but we return to the issue of the proper
choice of $f$ in \S\ref{discuss}. 

The emergent planetary spectra are calculated at 5000 wavelength
points distributed logarithmically from 0.3 to 300 microns. The temperature-pressure
profile in flux equilibirum is derived from $\sim$$10^{-6}$ to $\sim$$10^3$ bars.  The spectral models
for the stellar primaries (given a specific spectral subtype) come from Kurucz (1994).  
The molecular and atomic opacities are taken from the opacity library described in
Burrows et al. (2001) and equilibrium compositions are derived using the updated thermochemical  
database of Burrows \& Sharp (1999). Though the atmosphere is 
close to an ideal gas, the $H/He$ equation of state 
of Saumon, Chabrier, \& Van Horn (1995) that incorporates 
non-ideal effects is used.

\section{Comparison of Theoretical Planet-Star Flux Ratios}
\label{comparison}

Figure \ref{fig:2} depicts our theoretical planet-star flux-density ratios
versus wavelength in the near- and mid-infrared for TrES-1 (magenta)
and HD209458b (green).  To derive these curves we have used the physical data
for the planets and their primaries described in 
\S\ref{data}.  Our baseline planet models have solar metallicity.  
The phase-averaged (Sudarsky, Burrows, \& Hubeny 2003)
flux ratios, but for $f=0.25$ (\S\ref{model}; Sudarsky, Burrows, \& Hubeny 2003; 
Burrows, Sudarsky, \& Hubeny 2004), are given and have not been 
shifted in any way.  Superposed are the new data
at 24 \mic for HD209458b (green) and at 4.5 \mic and 8.0 \mic for TrES-1 (gold).
The vertical error bars are the quoted 1-$\sigma$ ranges and the horizontal
bars indicate the widths of the corresponding IRAC and MIPS bands.
Also included in yellow are the four theoretical IRAC band-averaged fluxes
for the TrES-1 model and in blue for the HD209458b model.
These band averages are derived using the published transmission functions
\footnote{See the IRAC web page at http://ssc.spitzer.caltech.edu/IRAC/},
divided by frequency to obtain the theoretical ratio of detected electrons.
Since the close-in EGP (this work) and stellar (Kurucz 1994)
spectra are so flat at 24 \mic, a transmission band-averaged 
point is not shown (or needed) there.

Varying \mp, \rp, and \rstar\ within the error bars alters the 
resulting planet-star flux-density ratios only slightly.  Similarly, and perhaps surprisingly,
adding Fe and forsterite clouds does not shift the predictions in the 
Spitzer bands by an appreciable amount.  Moreover, despite
the more than a factor of two difference in the stellar flux at the planet, the predictions
for the planet-star ratios for two such disparate close-in EGPs as TrES-1 and HD209458b are not 
very different.  For TrES-1, raising or lowering the metallicity by a factor of three 
changes the flux ratios by less than $\sim$10\%.  
Interestingly, however, changes in $f$ and, by inference, day-night
atmospheric differences and phase function effects can result in
25\% to 50\% deviations in the flux ratios of which one 
should take note.

Comparing our baseline models with data on Fig. \ref{fig:2}, we can deduce several
interesting things.  First, the 24-\mic data point is close to the predicted
value, though for all three data points the theory slightly underestimates the data by 
a factor of $\sim$1.5-1.8, with the largest discrepancy being for the TrES-1 band at 8.0 \mic.  
For HD209458b at 24 \mic this deviation is only $\sim 1 \sigma$.
The high absolute values of the flux ratios imply that the close-in EGP atmospheres are
indeed at high temperature, predicted for TrES-1 and HD209459b 
to be $\sim$1500 K and $\sim$1600 K, respectively, at a Rosseland 
depth of $\sim$1.  Both atmospheres clearly span the temperature range 1000-2000 K.
For TrES-1 at 8 \mic, theory yields a brightness 
temperature (temperature at $\tau_{\lambda} = 2/3$) near
800-900 K, slightly lower than the $\sim$1100 K crudely inferred from the 
data.  At 4.5 \mic, the theoretical brightness temperature 
of TrES-1 is $\sim$750-900 K (using $f=0.25$),
again slightly lower than the data might imply.  However, 
care must be taken in estimating temperatures of any sort, and we
will not, due to the hazards of extracting an ``effective" or ``equilibrium"
temperature from these data, say much more about them.  We do note, however, that
given the radius of HD209458b and its general flux level, we expect that its
bolometric luminosity is above $2\times 10^{-5}$ L$_{\odot}$.  This is approaching the 
luminosity of a {\it star} $\sim$100$\times$ the mass of HD209458b at the edge of 
the hydrogen-burning main sequence (Burrows et al. 2001).

In Fig. \ref{fig:2}, there is a hint
of the presence of H$_2$O, since it is expected to suppress flux between 4 \mic
and 10 \mic. This is shortward of the predicted 10-\mic peak in planet-star flux-density ratio,
which is due to water's relative abundance and the strength of its
absorption bands in that wavelength range. Figure \ref{fig:3} demonstrates this by
comparing H$_2$O, CH$_4$, and CO opacities per molecule from 0.5 to 10 microns.
Without H$_2$O, the fluxes in the IRAC bands would be much higher than the fluxes
in the mid-infrared. Hence, a comparision of the TrES-1 and HD209458b data at 4.5/8.0 \mic
and 24 \mic suggests, but does not prove, the presence of water.  As Fig. 
\ref{fig:3} implies, seeing the expected slope between the 5.8-\mic 
and 8.0-\mic bands and the rise from 4.5 \mic to 3.6 \mic would be more 
revealing in this regard and is a prediction of our theory.
Furthermore, the relative strength of 24-\mic MIPS flux ratio in 
comparison with the 3.6-\mic, 4.5-\mic, and 5.8-\mic
IRAC channel ratios is another prediction of the models, as is the
closeness of the 8.0-\mic and 24-\mic ratios.  The latter seems borne out by 
the data, though these data are for two different objects.

Nevertheless, the models have difficulty fitting the depth of the 4.5-\mic
feature in TrES-1.  This feature coincides with the strong CO
absorption predicted to be a signature of hot EGP atmospheres (Fig. \ref{fig:3}; Sudarsky, Burrows,
\& Hubeny 2003; Burrows , Sudarsky, \& Hubeny 2004; Burrows 2005), but is 
shallower than expected and $\sim$2-$\sigma$ discrepant. 
The data are in fact the band-averaged flux-density
ratios of the detected electrons.  As such, the larger fluxes on either side
of the trough theoretically centered close to 4.5 \mic contribute
planet flux to the detected band.  As a result, the yellow dot that represents this integrated
band contribution is a weak function of CO abundance.  
In fact, a CO abundance 100$\times$ larger than expected in chemical equilibrium lowers this flux ratio
at 4.5 \mic by only $\sim$25\%. Therefore, while the 4.5-\mic data 
point for TrES-1 implies that CO has been detected, the exact fit is problematic.  
More data and further attention to calibration are called 
for.  However, as we have indicated and discuss 
in \S\ref{discuss}, the contrast between theory and measurement at all data points 
may be a signature of day-night infrared flux asymmetry.  
The close-in planets may not be radiating heat energy
isotropically (as is assumed when using $f = 0.25$), 
a not-unexpected result (Guillot \& Showman 2002).

\section{Discussion}
\label{discuss}

\begin{figure}
\epsscale{1.00}
\vspace*{-0.7in}
\plotone{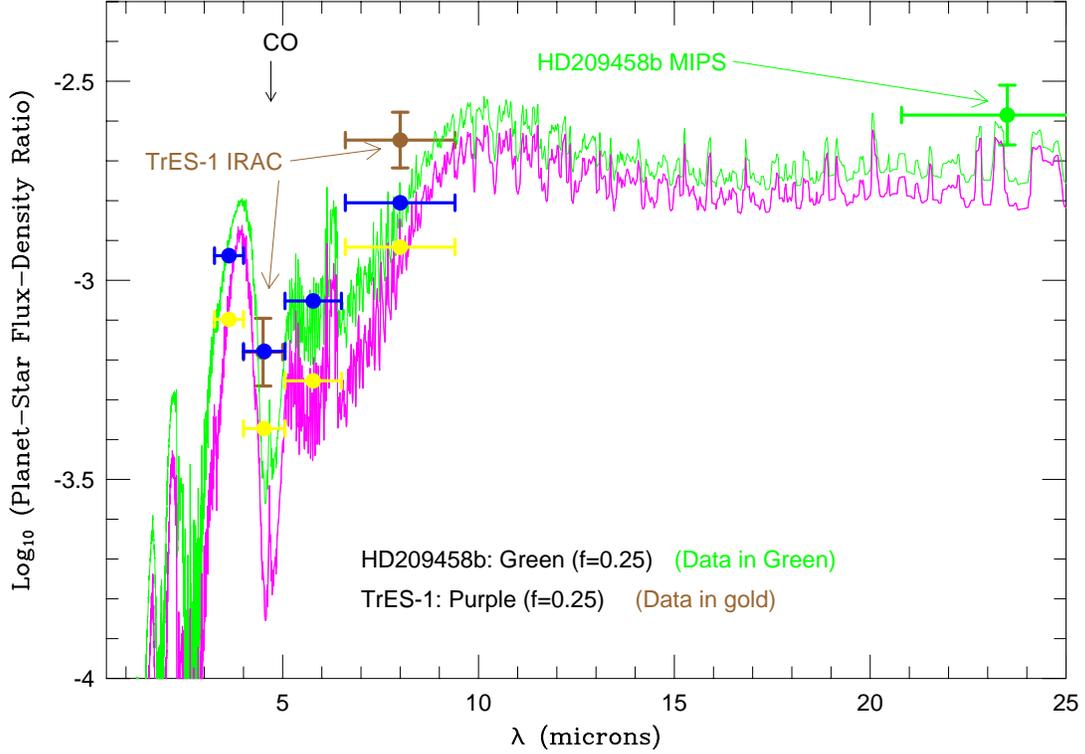}
\vspace*{-0.2in}
\caption{
The logarithm base ten of the planet-to-star flux-density ratio as a function
of wavelength ($\lambda$, in microns) for our baseline models of TrES-1 and
HD209458b (for $f$ = 0.25).  The model for TrES-1 is purple and that for
HD209458b is green. Superposed are the secondary eclipse data: the gold dots with 
corresponding error bars are the TrES-1 Spitzer/IRAC data from Charbonneau et al. (2005), while
the green dot with error bars is the HD209458b Spitzer/MIPS 24-\mic datum
from Deming et al. (2005).  Also included are the band-averaged 
detected-electron/``flux" ratios for the TrES-1 (yellow) and 
HD209458b (blue) models in the four IRAC bands.  Note that coincidently 
the blue dot at 4.5 \mic overlaps the gold TRES-1 data point.
The position of the strong CO absorption feature 
at $\sim$4.67 \mic is indicated and clearly coincides 
with the $\sim$4.5 \mic IRAC band flux. See text for a discussion and details.
\label{fig:2}}
\end{figure}

\begin{figure}
\epsscale{1.00}
\vspace*{-0.7in}
\plotone{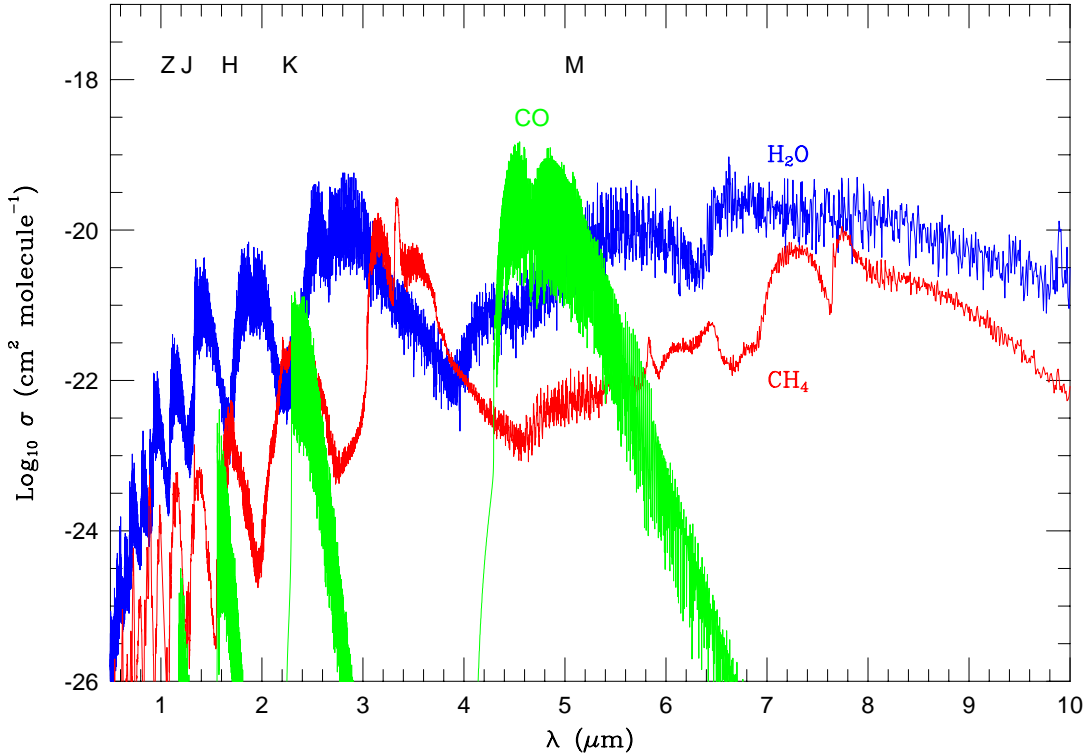}
\vspace*{-0.2in}
\caption{
The logarithm base ten of the absorption cross section per molecule
(in cm$^2$) versus wavelength (in microns) from 0.5 \mic to 10 \mic
for water (H$_2$O), carbon monoxide (CO), and methane (CH$_4$).  
These numbers are not weighted by abundance.  A comparison with Fig. \ref{fig:2}
clearly shows that CO is probably present in TrES-1 and that water helps shape the 
planet-star flux-density ratio in the 4 to 9 micron spectral region.
Shown also are the positions of various standard photometric bands ($Z, J, H, K, M$).
This figure, in conjunction with Fig. \ref{fig:2}, demonstrates that if CH$_4$ is present in abundance in 
either HD209458b or TrES-1 then the 3.6-\mic IRAC band will test this.  However,
our expectation for these close-in EGPs is that CH$_4$ will not be in evidence.   
When data for all four IRAC bands are available,
we should be able to verify definitively the presence 
of water vapor in the atmosphere of an extrasolar giant planet,
though by the absolute level at 8.0 \mic of the TrES-1 flux in Fig. \ref{fig:2}
the presence of water already seems likely.
See text for further discussion.
\label{fig:3}}
\end{figure}

We predict that both the 3.6-\mic and the 5.8-\mic flux ratios
will be higher than the 4.5-\mic ratio by at least 50\% and that the
pattern of the yellow and blue dots on Fig. \ref{fig:2} will be realized.
We also predict that the 24-\mic flux ratio for TrES-1 will be similar
to that seen for HD209458b (Deming et al. 2005).  However, while 
the close correspondence of the measured and theoretical fluxes
depicted on Fig. \ref{fig:2} is striking, it is not perfect.  
What could explain the differences?  One major uncertainty is the 
day-night atmospheric profile difference.  HD209458b and TrES-1 are
close enough to their primaries to be in synchronous rotation.  Therefore, they show the
same hemisphere to the star at all times.  It is the zonal winds, atmospheric circulation currents,
and jet streams (Menou et al.2002; Guillot \& Showman 2002; Showman \& Guillot 2002;
Cho et al. 2003; Burkert et al. 2005) that advect heat from the day to the night
sides, thereby affecting the atmospheric temperature structure as a function
of longitude.  How much of the stellar radiation goes to heating the day
side (visible just before and after the secondary eclipse) and how much
is transported by mass motion away from the day side to heat the night side?
This question remains unresolved, but directly impinges upon the 
planet-star flux-density ratios in the infrared measured during secondary eclipses. 

The $f$ factor we use for our fiducial model (0.25) is tailored
to distribute heat over the entire planet, on both the day and the night sides.  
The three data points depicted in Fig. \ref{fig:2} are all factors of $\sim$1.5-1.8 above
the $f=0.25$ theory curves.  This implies that the planet's radiation
is predominantly radiated by the day side and that its emissions are
forward-beamed.  This is not entirely unexpected in the optical, but
seems to be the case in the mid-infrared as well.  
The implication of the new secondary eclipse data at 24 \mic and 8 \mic may be that we 
are seeing indirect signs of an asymmetry in the day-night
heating and temperature profiles, with the day side hotter than
the night side by at least 500 K.  This estimate is based on the
mid-infrared ``excesses" seen in Fig. \ref{fig:2}, 
on the possible model variations, and on the measurement errors.  
However, the confirmation of such a conclusion awaits more detailed
multi-dimensional general circulation models with correct transport,
next-generation models of the wavelength-dependent phase functions 
of close-in EGPs (Sudarsky, Burrows, Hubeny, \& Li 2005), and, 
most importantly, additional data.

It has long been suggested, and recently articulated (Cooper \& Showman 2005),
that due to winds the sub-stellar point of an irradiated close-in EGP may not
be the hottest spot.  Advection would introduce a lag even for circular orbits between the
planet's ephemeris and its light curve (the light curve would lead).
Cooper \& Showman (2005) estimate that the lead could be as much as 60$^{\circ}$
and could amount to a 20\% brightness shift in the value at superior conjunction.
A 20\% decrement is not enough to close the modest apparent gap between our theory
and the 24-\mic data for HD209458b, and is not of the correct sign. 
However, the concept of a shift in the light curve deserves further scrutiny.   

Charbonneau et al. (2005) estimate a Bond albedo ($A$) for TrES-1 of 0.31$\pm$0.14.
However, one should be very cautious using these new data
to infer temperatures and reflection coefficients.  Not only is the reradiation
not expected to be isotropic off the planet, but the atmospheres are not black bodies.
While one can distinguish hot atmospheres (1000-2000 K) from cooler atmospheres
(500-1000 K), the data and theory are not yet adequate to allow these TrES-1
data to strongly constrain $A$.  However, if $A$ were in the 0.31$\pm$0.14
range, this would imply that there is a cloud of non-trivial optical depth
in the upper layers of TrES-1, putting it into the ``Class V" category
of EGPs, rather than the ``Class IV" category (Sudarsky, 
Burrows, \& Pinto 2000).  The latter, due to strong
absorption bands and little Rayleigh scattering, have very low Bond albedos 
(below 0.05).  Our models for TrES-1 would favor the low-albedo Class IV 
category for TrES-1 and the higher-albedo Class V category for HD209458b,
but we feel it is premature to conclude anything definitive about albedos
at this stage (other than that they can not be very high and that the close-in EGPs 
can not be highly reflective).

In this paper, we have calculated planet-star flux-density ratios
versus wavelength, focussing on the near- and mid-infrared out to
25 microns and the irradiated close-in extrasolar giant planets
TrES-1 and HD209458b, and have compared our theory with the recent 
secondary eclipse data from Charbonneau et al. (2005) and 
Deming et al. (2005). We have inferred the presence of carbon
monoxide, and perhaps water, in the atmosphere of TrES-1 and have determined
that both atmospheres are hot ($\sim$1000-2000 K).  We have explored the effects of
varying the metallicity, \mp, \rp, and \rstar, the latter three 
within the stated error bars, and find our predictions 
for the planet-star flux-density ratios to be robust. However, 
more work is required to understand the apparent shallowness of the TrES-1 
4.5-\mic trough.  (Non-LTE and non-equilibrium effects in the chemistry may be worth exploring, as well
as the consequences of possible stratospheric temperature inversions (Smith \& Hunten 1990).) 
However, we suggest that the data (particularly at 8 and 24 microns) 
indicate we are beginning to constrain the degree of anisotropy
in the temperature profile of a close-in EGP (the day-night contrast) and in the angular
dependence of its emission.  Our preliminary conclusion is that the slight systematic
differences seen in Fig. \ref{fig:2} between the 
phase-averaged theory and all three new measurements may be explained in part
by an infrared-brighter, hotter day side.  This raises 
the intriguing possibility that additional and more
precise secondary eclipse data for these transiting EGPs could
shed light on their global meteorology.  Be that as it may, 
these secondary eclipse data are opening a new chapter in the accelerating 
study of extrasolar planets and emphasize that knowledge of a unique  
character is generated when new capabilities emerge.

\acknowledgments

We thank Christopher Sharp, Bill Hubbard, 
and Drew Milsom for useful discussions
during the course of this investigation and Dave Charbonneau for 
an advanced look at the TrES-1 secondary eclipse data. This study 
was supported in part by NASA grants NNG04GL22G and NAG5-13775.
This material is based upon work supported by the National Aeronautics and
Space Administration through the NASA Astrobiology Institute under
Cooperative Agreement No. CAN-02-OSS-02 issued through the Office of Space
Science.

{}

\clearpage

% figure 1

% figure 2

% figure 3

\end{document}